\documentclass[aps,prl,twocolumn,english,superscriptaddress,showpacs]{revtex4-1}
\usepackage[T1]{fontenc}
\usepackage[latin9]{inputenc}
\usepackage{babel}
\usepackage{amsmath}
\usepackage{amssymb}
\usepackage{graphicx}
\usepackage{wasysym}
\usepackage{graphicx}
\usepackage{xcolor}

\makeatletter

\usepackage{subfigure}

\makeatother

\usepackage{babel}
\begin{document}

\title{Ab initio calculation of a Pb single layer on a Si substrate: two-dimensionality
and superconductivity}

\author{A. Linscheid} 
\affiliation{Max-Planck-Institut für Mikrostrukturphysik, Weinberg 2, D-06120 Halle (Germany)}

\author{A. Sanna} 
\affiliation{Max-Planck-Institut für Mikrostrukturphysik, Weinberg 2, D-06120 Halle (Germany)}

\author{E. K. U. Gross}
\affiliation{Max-Planck-Institut für Mikrostrukturphysik, Weinberg 2, D-06120 Halle (Germany)}

\begin{abstract}
\noindent We report on first principles calculations of superconductivity
in a single layer of lead on a silicon substrate including a full
treatment of phononic and RPA screened coulomb interactions within
the parameter free framework of Density Functional Theory for superconductors.
A thorough investigation shows that several approximations that are commonly
valid in bulk systems fail in this constrained 2D geometry. The calculated
critical temperature turns out to be much higher than the experimental
value of $1.86\mbox{K}$. We argue that the only plausible explanation
for the experimental $T_{c}$ suppression is the onset of fluctuations
of the superconducting order parameter. 
\end{abstract}

\maketitle
Nature shows a clear correlation between superconductivity
and dimensionality as all superconductors with a high critical temperature
($T{}_{c}$), cuprates, pnictides and MgB$_2$, have sharp
two-dimensional properties. Understanding this connection is among 
the most important targets in contemporary solid state research. It
is likely that relevant physical mechanisms work differently in reduced
dimensionality and that approximations and theoretical methods developed
through the experience accumulated on three-dimensional systems have
to be modified for constrained geometries. Moreover fluctuation instabilities
of the order parameter may play an important role\cite{MerminWagner}.

The conclusive test to check the theoretical understanding is to perform
\textit{ab initio} calculations and compare directly with experiments.
Currently, such a test cannot be done for pnictides and cuprates as
the pairing mechanism is still under debate. However, it can be done
for phononic superconductors.

In this work we present the results of this type of analysis for lead,
as this phononic superconductor is correctly described in bulk by
ab-initio methods\cite{Floris,Bersier} and is experimentally realized in
the two-dimensional limit by deposition on a silicon 111 substrate\cite{Zhang,Yamada,Stepniak,Horikoshi}.

\begin{figure}[h]
\includegraphics[width=0.95\columnwidth]{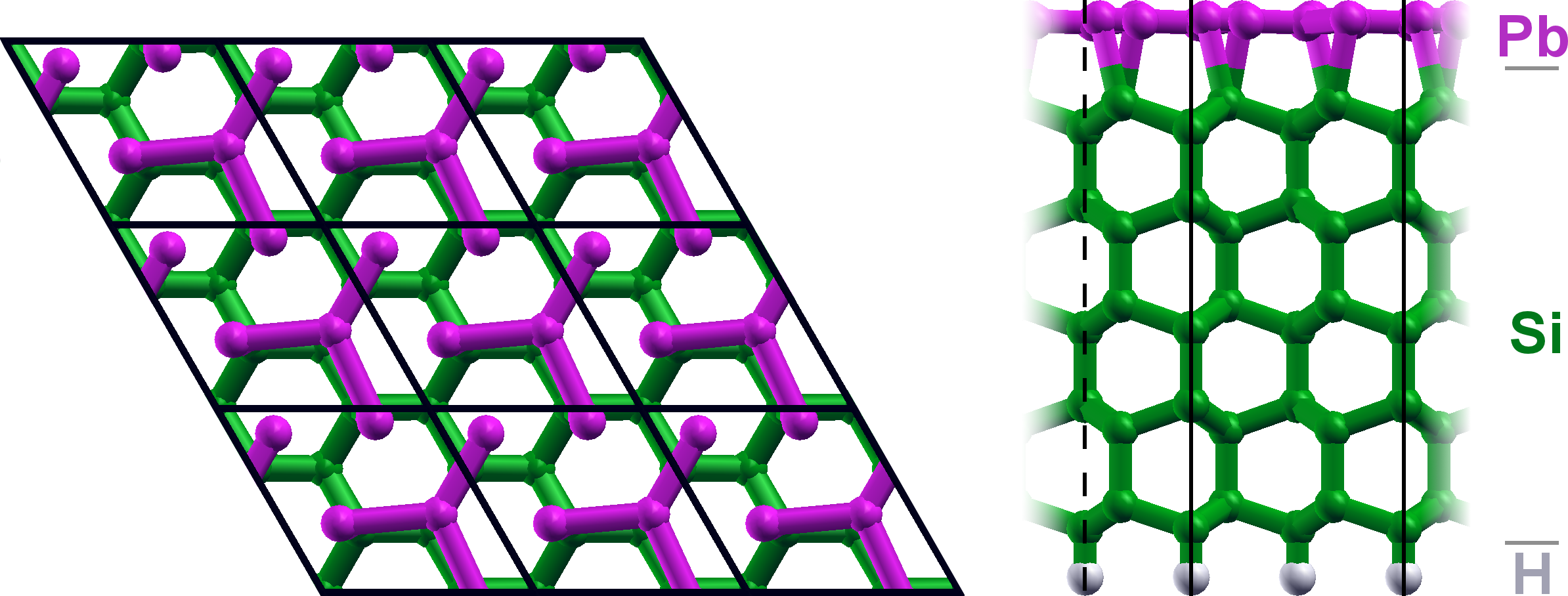}
\caption{(color online) SIC configuration of Pb on the Si (111) substrate.
On the left we present the top view, and on the right the side view.
Black lines mark the simulated unit cell.}
\label{fig:structure} 
\end{figure}

This Si-Pb system is constructed as shown in Fig.~\ref{fig:structure}.
We model the Si substrate by a 111 oriented slab, which is passivated
on the opposite side of the lead surface using hydrogen\cite{Noffsinger,Cudazzo}. A relatively
large width of five Si-bilayers is chosen in order to reduce spurious
size effects of the substrate on the Pb layer. For the same reason
we constrain the hexagonal (xy) Si unit cell to its bulk size. Lead
is placed in the so-called striped incommensurate (SIC) configuration.
Since we work with periodic boundary conditions, a vacuum of $\sim$8~\AA{}\ separates
the periodic replica of the system. Within these constraints a full
relaxation is performed. Relaxations, electronic structure, phonons
and electron-phonon interactions have been calculated within Kohn-Sham
(KS) density functional theory (DFT)\cite{comp_details}. The calculated
electron phonon coupling strength results in $\lambda=0.78$. If we
use the McMillan formula\cite{McMillan} with a standard value for
the parameter $\mu^{\ast}=0.10$ we obtain an estimation for the critical
temperature of 1.98K. This is in very good agreement\cite{comment_cohen} with the experimental
$T_{c}$ of 1.86K\cite{Zhang}. With this result one
has to conclude that superconductivity in this 2D limit can be understood
from the electronic coupling alone and no fluctuations are necessary
to explain the physics of this system.

Is this really the end of the story? Is it correct to assume the validity
approximations known to work well in bulk superconductors also for
this low dimensional system? The answer is no, and to show this we
proceed to deeper investigation.

In order to avoid any adjustable parameter (as the above named $\mu^{\ast}$)
we use density functional theory for superconductors\cite{OGK,Marques,Lueders,MgB2,LiKAl,CaC6,Floris,H,AkashiNitrides,AkashiC60,plasmonicSCDFT} (SCDFT), where electronic and phononic couplings are included on the
same footing.

\begin{figure}[h]
\includegraphics[width=0.4\textwidth]{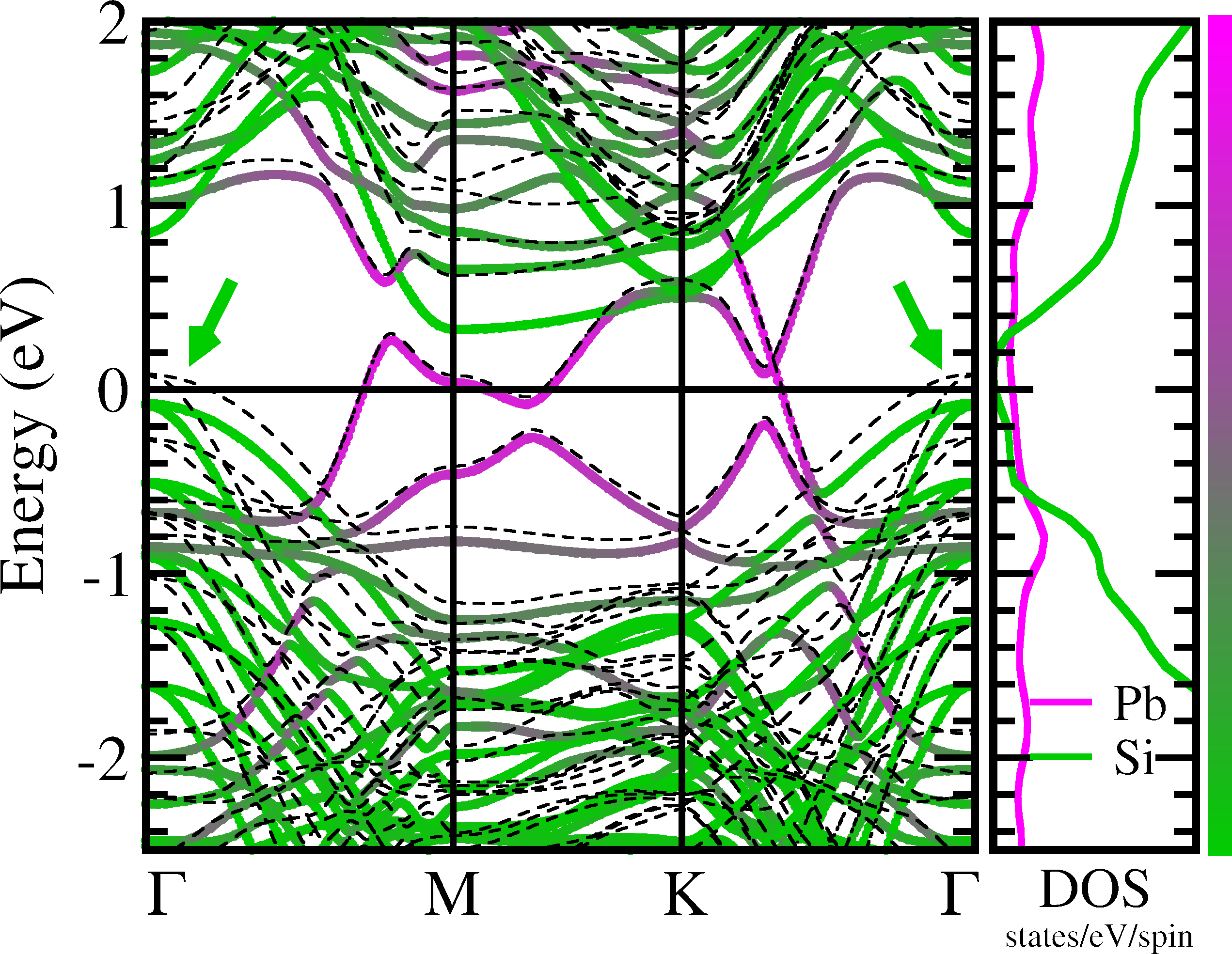}

\caption{(color online) Band Structure (left) and density of states (right)
near the Fermi Energy. Thick lines correspond to a phosphorus doped
system (1 part per 224), and the color scale corresponds to the
projection on the Pb states. Dashed lines are the bands in the undoped
system. The P doping has a negligible effect on Pb bands while it
completely fills the silicon hole pockets (indicated by green arrows). }

\label{fig:bands} 
\end{figure}

\textit{Electronic and phononic properties -} A very relevant property
in the electronic structure (Fig.~\ref{fig:bands}) is the presence
of both Pb and substrate metallic bands. This means that Pb deposition
acts as a dopand to the Si substrate which develops a surface metallic
region. This metallic region fades away within a few layers. The presence
of this additional metallic band is relevant for two reasons. First
it may provide a contribution to the electron-phonon coupling
and, second, it may stabilize fluctuations of the order parameter
of the superconducting phase by effectively enhancing the three-dimensionality
of the condensate. These Si metallic bands can be removed by using
an n-doped substrate. We explicitly consider this case by substituting
one Si atom (in the deep bulk) with a virtual mixture of P and Si,
corresponding to a doping of 1 part per 240 Si. Doping has a small
effect on the filling level of the Pb bands, but completely saturates
the Si- hole pockets (see Fig.~\ref{fig:bands}). This doped system
is experimentally realized \cite{Zhang} and allows for direct comparison
with results obtained in this study.

\begin{table}
\begin{tabular}{c|ccccccc}
 & $\lambda^{Pb,Pb}$  & $\lambda^{Pb,Si}$  & $\lambda^{Si,Si}$  & $\lambda_{av}$  & max$\left[\lambda_{i}\right]$  & N$_{Pb}(0)$  & N$_{Si}(0)$ \\
\hline 
undoped & 0.95  & 0.13  & 0.06  & 0.78  & 0.98  & 0.97  & 0.60 \\
doped  & 1.03  & 0.00  & 0.00  & 1.03  & 1.03  & 1.07  & 0.00 \\
\end{tabular}\caption{Electron phonon coupling coefficients. $\lambda^{i,j}$ is the Fermi
Surface sheet resolved coupling matrix. $\lambda_{av}=\frac{1}{N(0)}\sum_{i,j}\lambda^{i,j}N_{i}(0)$
is the average electron phonon coupling where $N{}_{i}(0)$ are the
Fermi surface resolved DOS and $N(0)$ is the total DOS. max$\left[\lambda_{i}\right]$
is the maximum eigenvalue of $\lambda$ that in BCS acts as the effective
pairing to determine the critical temperature\cite{Suhl}. }

\label{tab:el-ph} 
\end{table}

The degree of two-dimensionality of the SC phase, i.e. how the condensate
extends into the substrate, is determined by the lead-substrate interaction.
We can distinguish three main effects that describe how the Pb surface
and the substrate are coupled: chemical hybridization, electron phonon
coupling and Coulomb interaction.

The \textit{chemical hybridization} between surface and substrate
states can be made visible by projecting the KS states on the Pb atomic
orbitals. This analysis shows that the KS states near the Fermi energy
are either located in the lead surface or inside the silicon bulk,
with no overlap (see Fig.~\ref{fig:bands}).

The \textit{electron-phonon coupling} is computed for the KS system
via linear response\cite{comp_details}. Phonons may generate pairing
between bulk and surface states. In Tab.~\ref{tab:el-ph} we report
the FS-resolved el-ph coupling\cite{footnotemultiband}. By considering
the average coupling and ignoring the energy dependence of density
of states and screened coulomb interactions (by approximating them
with the value at the Fermi energy) we have a formal equivalence of
SCDFT with the McMillan method. The resulting critical temperature
of lead on the undoped substrate is $T_{c}=2.01\mbox{K}$ and $T_{c}$
rises to $2.74\mbox{K}$ for the doped Si substrate. The difference
in critical temperatures between the doped and undoped system is caused
by the fact that the undoped material has an mean coupling which is
much weaker than the lead-lead intra-surface coupling alone. This
implies that the isotropic approximation is unjustified and leads
to an underestimation of $T_{c}$. Multiband-superconductivity must
be explicitly accounted for as in the well known case of $\text{MgB}_{2}$.

Moreover the electrons are subject to a \textit{screened Coulomb}
scattering which we treat within the RPA\cite{cac6_sust,FlorisSannaMgB2}. This kind
of interaction in bulk materials is often overlooked, since, acting
both as a repulsive (directly) and attractive interaction (via Coulomb
renormalization mechanism\cite{Morel_Anderson,ScalapinoSchriefferWilkins,AllenMitrovic,Schrieffer})
it appears very often to be largely material independent. This shows
up in Eliashberg based methods\cite{Eliashberg,McMillan} in the well-known
rule of thumb to take $\mu^{\ast}\sim0.1$. A crucial advantage of SCDFT is that via the matrix elements of the RPA-screened Coulomb interaction the Coulomb renormalization effect is explicitly calculated, making the use of empirical parameters like $\mu^*$ obsolete. 
 A metallic layer on a semiconducting substrate is conceptually different from a bulk
in that, due to the lower dimensionality, there is a reduced phase
space for low energy Coulomb scattering, that is repulsive for Cooper
pairing (in s-wave), while the space for high energy scattering is
not restricted, owing to the presence of the substrate. Therefore
the Coulomb renormalization is unusually large in this type of system.

\begin{table}
\begin{tabular}{c|ccc|c}
 & T$_{c}$  & $\Delta^{Pb}(0)$  & $\Delta^{Si}(0)$  & $T_{c}^{*}$ \\
\hline 
undoped & 3.42  & 0.71  & 0.32  & 2.01 \\
doped  & 3.54  & 0.74  & ---  & 2.74 \\
\end{tabular}\caption{ Calculated critical temperatures, T$_{c}$ (in K), within SCDFT and
superconducting gap, $\Delta$ (in meV), on the Lead and Si Fermi
surfaces. T$_{c}^{*}$ is the critical temperature estimated using
an average coupling on the Fermi surfaces, ignoring the energy dependence
of dos and screened coulomb interactions (corresponding to a $\mu^{*}$
like approximation). }

\label{tab:tc} 
\end{table}

\textit{Discussion -} The computed critical temperature for the undoped(doped)
system as given in Tab.~\ref{tab:tc} is 3.42(3.54)K. 

We have then shown that, releasing several unjustified approximations,
the estimated critical temperature of 3.54K (doped system) is far
too high as compared to the values experimentally observed 1.86K\cite{Zhang}, 1.5K\cite{Stepniak} 
and 1.1K\cite{Yamada}.

What is the source of this mismatch? To answer this question we have
to carefully investigate the effects not considered in the above analysis 
 and their possible influence on superconductivity. I) We have assumed the RPA represent the screened
Coulomb interaction. This is reliable in the high-density limit when
screening is good. Therefore the Pb layer is expected to be well described.
The approximation may be less accurate for the silicon hole band, since these
states have a low density and, thus, will be poorly screened. However,
the strong Coulomb repulsion will prevent a significant contribution
to superconductivity, therefore this inaccuracy cannot affect the
estimated $T_{c}$ significantly. Surely not for the doped system
where these bands do not even cross the Fermi level. II) In general,
when computing the electron phonon pairing, 
vertex corrections can be safely dropped, due to Migdal's theorem\cite{Eliashberg,Migdal}.
The shape of the Si hole pocket band might invalidate this conclusion. However, this cannot
have a significant influence on the calculation of superconductivity
in this system since, as discussed above, this band effectively does
not take part in the condensation.
Migdal's theorem is also not applicable in the small q limit. 
This does not affect the estimation of the phononic pairing,
due to the small fraction of the Brillouin zone in which the problem occurs.
Nevertheless we have to keep in mind that the low q physics are not
correctly described under this assumption.
III) In our calculations we do not include spin-orbit coupling effects.
These have been shown to be relevant both for bulk lead\cite{Heid,Sklyadneva_bulk} and lead multilayers\cite{Sklyadneva_layers}. However the effect systematically increases the coupling strenght, therefore it can not explain our overestimation of T$_c$. And actually its inclusion would lead to an even higher critical temperature.
IV) In our work we consider only
a statically screened Coulomb interaction. The result of dynamic (plasmonic)
effects could lead to important modifications of the dielectric screening
in the case of low energy surface plasmons. However, as first pointed out by Takada, this effect is
known to give a positive contribution to superconductivity (enhancement
of coulomb renormalization by the plasmonic peak\cite{Takada,RietschelSham,plasmonicSCDFT} ). Therefore, if relevant,
it would lead to a higher estimate of $T_{c}$. V) Another questionable
approximation is the use of the LDA in the low dimensional limit.
This issue has been investigated in detail by Pollack and Perdew\cite{PollackPerdew}
showing that LDA performs well as soon as the ratio between the layer
thickness and the $r_{s}$ coefficient of the gas is $\approx2$.
In our case this ratio can be estimated to be of the order of $5$
and we expect the LDA to perform as reliably as usual. VI) Due to the
poor metal-substrate coupling, the calculated single particle excitation
spectrum of Si presents a fundamental gap that is about one half of
the observed gap in bulk silicon. This may lead to an overestimation
in the Coulomb renormalization, and then in an overestimation of $T_{c}$.
We have therefore accounted for this effect in our calculations by
including a scissor correction on the Si bands and the resulting effect
on $T_{c}$ correction is $<$~0.1K.

We believe that we have considered all relevant electronic
pairing effects. In the bulk limit the critical temperature in SCDFT, using the same approximations as for the slab is 6.3 K that compares well with the experimental value of 7.2 K.

The only mechanism that is not included in our simulations and that, according
to model calculations, is strongly suspected to suppress superconductivity,
is the onset of fluctuations in the order parameter. While this could
be in principle captured in SCDFT, the presence of infrared collective
excitations of the order parameter is not accounted for in the present
functionals. Owing to Mermin-Wagner's theorem\cite{MerminWagner,Hoenberg}
these fluctuations completely forbid superconductivity in a strictly two dimensional system.
In 3D systems of constrained geometry (such as surfaces) model calculations
show that these fluctuations may still be relevant in the limit in
which the thickness is of the atomic scale and the in-plane dimension
of the system is macroscopic\cite{JasnowFisher}. Due to the strong
confinement of the SC phase to the lead layer, as is clearly seen
in the real space structure of the order parameter of Fig.~\ref{fig:structure},
\begin{figure}
\begin{minipage}[t][1\totalheight][c]{1\columnwidth}
\begin{minipage}[c]{0.5\columnwidth}
\includegraphics[width=1\textwidth]{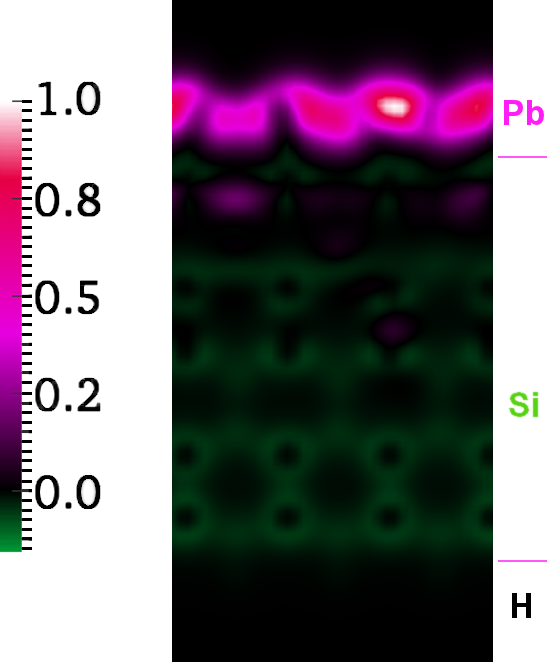}
\end{minipage}
\end{minipage}

\caption{(color online) Real-space structure of the SC order parameter $\chi(\mathbf{R},\mathbf{0})$
normalized to its maximal value of $0.0002765$. As it can be interpreted
as the wave function of condensed pairs the confinement of the SC
phase to the Pb layer is clearly visible. The dark blue in the substrate
indicates that it takes part in the Coulomb renormalization and thus
reflects a proximity effect that extends throughout the whole substance.}
\end{figure}
 one would expect to be in a regime where these fluctuation effects
of the superconducting order parameter are relevant.
While, as mentioned, neglecting vertex corrections has probably little effect on the phononic el-el coupling,
effective interactions in the superconducting Nambu channel in the sense of the fluctuation propagator \cite{Larkin} 
can be very important.
The disagreement between the calculated and experimental
critical temperature then strongly suggests that $T{}_{c}$ is experimentally
limited by the fluctuation regime. The superconducting phase rapidly
stabilizes with an increasing number of Pb layers\cite{Qin,Brun,Eom} strengthening
this conclusion.

To summarize, we report a first-principles calculation of
the superconducting ground state of a single lead layer deposited
on a Si (111) substrate. We account for phonon mediated (via linear
response DFT) and screened coulomb pairing (RPA) within the parameter
free framework of Superconducting Density Functional Theory. We have
shown that the isotropic approximation is not valid in this surface
configuration, and in particular that the isotropic $\mu^{\ast}$ approximation
used for bulk superconductivity leads to a large underestimation of
the critical temperature. Our calculations predict a critical temperature
about 80\% larger than observed in experiment. Our analysis strongly
suggests that this mismatch is attributed to the onset of long wavelength
phase fluctuations of the superconducting order parameter.


\begin{thebibliography}{10}

\bibitem{MerminWagner} N.D.~Mermin and H.~Wagner, Phys. Rev. Lett.
\textbf{17} 1133, (1966).

\bibitem{Floris} A.~Floris, A.~Sanna, S.~Massidda and E.K.U.~Gross,
Phys. Rev. B \textbf{75}, 054508 (2007).

\bibitem{Bersier} C.~Bersier \textit{et. Al}, J. Phys.: Cond. Mat. {\bf 21}, 164209 (2009). 

\bibitem{Zhang} T.~Zhang \textit{et. Al}, Nat. Phys. \textbf{6},
104 (2010).

\bibitem{Yamada}
M.~Yamada, T.~Hirahara and S.~Hasegawa, Phys.~Rev.~Lett. {\bf 110}, 237001 (2013).

\bibitem{Stepniak}
A.~Stepniak {\it et al.}, Surf. Interface Anal. {\bf 5516} (2014).

\bibitem{Horikoshi}
K.~Horikoshi, X.~Tong, T.~Nagao and S. Hasegawa, Phys. Rev. B {\bf 60}, 13287 (1999).

\bibitem{Noffsinger} J.~Noffsinger and M.L.~Cohen, Solid.~State.~Comm. \textbf{151}, 421 (2011).

\bibitem{Cudazzo} P.~Cudazzo, G.~Profeta and A.~Continenza, Surf.~Sci. \textbf{602} 747 (2008).

\bibitem{comp_details} All normal state calculations have been done
in the local density approximation (LDA), for the exchange correlation
functional\cite{PZ}, and by using norm conserving pseudo-potentials
to account for core electronic states. A cutoff of 80 Ry has been
set in the plane wave expansion of the KS states and the Brillouin
zone has been sampled with a mesh of 12$\times$12$\times$1 k-points.
Phonons and electron phonon coupling have been calculated within linear
response DFT\cite{Baroni}, as implemented in the ESPRESSO package\cite{espresso}.

\bibitem{PZ} J.P.~Perdew and A.~Zunger, Phys.
Rev. B \textbf{23}, 5048 (1981).

\bibitem{Baroni} S.~Baroni, S.~de Gironcoli, A.~Dal Corso, P.~Giannozzi,
Rev. Mod. Phys. \textbf{73}, 515 (2001).

\bibitem{espresso} P. Giannozzi {\textit{e}t al.}, J. Phys.:Condens.
Matter \textbf{21} 395502 (2009), http://www.quantum-espresso.org/

\bibitem{McMillan}W.~L.~McMillan, Phys. Rev. {\bf 167}, 331 (1968).

\bibitem{comment_cohen}A similar agreement has been obtained before
by Noffsinger and Cohen\cite{Noffsinger}. We work in a slightly different
approximation, beyond using a larger Si bulk, we explicitly include
silicon phonon modes in our calculations.

\bibitem{OGK} L.N.~Oliveira, E.K.U.~Gross and W.~Kohn, Phys. Rev.
Lett. \textbf{60}, 2430 (1988).

\bibitem{Lueders} M.~L\"uders \textit{et al.}, Phys. Rev. B 72, 024545
(2005).

\bibitem{Marques} M.A.L.~Marques \textit{et al.}, Phys. Rev. B 72,
024546 (2005).

\bibitem{MgB2}
A.~Floris \textit{et al.}, Phys.~Rev.~Lett.  {\bf 94}, 037004 (2005).

\bibitem{LiKAl}
G.~Profeta {\it et al.}, Phys.~Rev.~Lett. {\bf 96},  047003  (2006).

\bibitem{CaC6}
A.~Sanna {\it et al.}, Phys.~Rev.~B {\bf 75},  020511(R)  (2007).

\bibitem{H}
P.~Cudazzo {\it et al.}, Phys.~Rev.~Lett. {\bf 100}, 257001 (2008).

\bibitem{plasmonicSCDFT}
R.~Akashi and R.~Arita, Phys.~Rev.~Lett. {\bf 111}, 057006 (2013).

\bibitem{AkashiNitrides}
R.~Akashi, K.~Nakamura, R.~Arita and M.~Imada, Phys.~Rev.~B {\bf 86}, 054513 (2012).

\bibitem{AkashiC60}
R.~Akashi and R.~Arita Phys.~Rev.~B {\bf 88}, 054510 (2013).

\bibitem{Suhl} H.~Suhl \textit{et. Al} Phys. Rev. Lett. \textbf{3},
552 (1959).

\bibitem{footnotemultiband} In the BCS framework, a multi-band system
is well characterized by its average coupling, a coupling matrix ($\lambda_{ij}$)
and its eigenvalues (see ref.~\onlinecite{Suhl}) for details).
T$_{c}$ is essentially determined by the highest eigenvalue of $\lambda_{ij}$,
if this number is close to the average coupling we say that the system
behaves as isotropic; if it is close to one of the diagonal elements
of the matrix then essentially the system is dominated by one sub-band,
and the others act hiding its coupling (this last situation is realized
in MgB$_2$).

\bibitem{cac6_sust}S.~Massidda \textit{et al.}, Supercond. Sci.
Technol. \textbf{22} 034006 (2009).

\bibitem{FlorisSannaMgB2}
A.~Floris \textit{et. Al} Physica C, {bf 456}, 45 (2007).

\bibitem{Morel_Anderson} P.~Morel and P.~W. Anderson, Phys. Rev.
\textbf{125}, 1263 (1962).

\bibitem{ScalapinoSchriefferWilkins} D.~J.~Scalapino, J.~R.~Schrieffer,
and J.~W. Wilkins, Phys.~Rev. \textbf{148}, 263 (1966).

\bibitem{AllenMitrovic} P.~B.~Allen and B.~Mitrovic, Solid State
Physics, edited by F.~Seitz (Academic Press, Inc., New York, 1982),
Vol.~37, p.\ 1.

\bibitem{Schrieffer} J.~R.~Schrieffer, Theory of Superconductivity,
Frontiers in Physics Vol.~\textbf{20} (Addison-Wesley, Reading, 1964).

\bibitem{Eliashberg} G.~M.~Éliashberg, Sov. Phys. JETP \textbf{11},
696 (1960).

\bibitem{Migdal} A.~B.~Migdal, Sov. Phys. JETP \textbf{7}, 996
(1958).

\bibitem{Heid}
R.~Heid, K.P.~Bohnen, I.Y.~Sklyadneva and E.~V.~Chulkov, Phys. Rev. B, {\bf 81} 174527 (2010).

\bibitem{Sklyadneva_bulk}
I.Y.~Sklyadneva, R.~Heid, P.M.~Echenique, K.P.~Bohnen and E.V.~Chulkov, Phys. Rev. B, {\bf 85}, 155115 (2012).

\bibitem{Sklyadneva_layers}
I.Y.~Sklyadneva, R.~Heid, K.P.~Bohnen, P.M.~Echenique and E.V.~Chulkov, Phys. Rev. B  {\bf 87}, 085440 (2013).

\bibitem{Takada} 
Y.~Takada, J. Phys. Soc. Jpn. {\bf 45}, 786 (1978).

\bibitem{RietschelSham} 
  H.~Rietschel and L.~J.~Sham, Phys. Rev. B {\bf  28}, 5100 (1983). 

\bibitem{PollackPerdew} L.~Pollack and J.P.~Perdew, J. Phys. Condens.
Matter \textbf{12} 1239 (2000).

\bibitem{Hoenberg} P.C.~Hohenberg, Phys. Rev. \textbf{158}, 383
(1967).

\bibitem{JasnowFisher} D.~Jasnow and M.E.~Fisher, Phys. Rev. B
\textbf{3}, 895 (1970).

A. Larkin and A. Varlamov, in Superconductivity, edited
by K. Bennemann and J. Ketterson (Springer Berlin Hei-
delberg, 2008) pp. 369¿458.

\bibitem{Larkin} A.~Larkin and A.~Varlamov, in Superconductivity,
edited by K.~Bennemann and J.~Ketterson (Springer Berlin Heidelberg, 2008) pp.~369-458.


\bibitem{Qin} S.~Qin, J.~Kim, Q.~Niu and C-K.~Shih, Science \textbf{324}
(2009).

\bibitem{Brun}
C.~Brun \textit{et al.}, Phys. Rev. Lett. {\bf 102}, 207002 (2009).

\bibitem{Eom} D.~Eom, S.~Qin, M-Y.~Chou and C.K.~Shih, Phys.
Rev. Lett. \textbf{96}, 027005 (2006).

\end{thebibliography}
\end{document}